\documentclass[aps,prl,twocolumn,longbibliography]{revtex4-2}

\usepackage{amssymb}
\usepackage{amsbsy}
\usepackage{amsmath}
\usepackage{graphicx}
\usepackage{graphics}
\usepackage{setspace}
\usepackage{array}
\usepackage{color}
\usepackage{fontenc}
\usepackage{textcomp}
\usepackage{bm}
\usepackage{float}
\usepackage[bookmarks=false,linkcolor=blue,urlcolor=blue,colorlinks,citecolor=blue]{hyperref}

\newcommand{\vex}[1]{\bm{\mathrm{#1}}}

\newcommand{\blue}[1]{{\color{blue}{#1}}}

\newcommand{\bsub}{\begin{subequations}}
\newcommand{\esub}{\end{subequations}}

\graphicspath{{../Figures/}}

\begin{document}
\title{Higher-Order Nodal Hinge States in Doped Superconducting Topological Insulator}
\author{Sayed Ali Akbar Ghorashi$^1$}\email{sayedaliakbar.ghorashi@stonybrook.edu}
\author{Jennifer Cano$^{1,2}$}
\author{Enrico Rossi$^3$}
\author{Taylor L. Hughes$^4$ }
\affiliation{$^1$Department of Physics and Astronomy, Stony Brook University, Stony Brook, New York 11794, USA}
\affiliation{$^2$Center for Computational Quantum Physics, Flatiron Institute, New York, New York 10010, USA}
\affiliation{$^3$ Department of Physics, William $\&$ Mary, Williamsburg, Virginia 23187, USA}
\affiliation{$^4$Department of Physics and Institute for Condensed Matter Theory,  University of Illinois at Urbana-Champaign, IL 61801, USA}

\date{\today}

\newcommand{\be}{\begin{equation}}
\newcommand{\ee}{\end{equation}}
\newcommand{\bea}{\begin{eqnarray}}
\newcommand{\eea}{\end{eqnarray}}
\newcommand{\h}{\hspace{0.30 cm}}
\newcommand{\vs}{\vspace{0.30 cm}}
\newcommand{\n}{\nonumber}
\begin{abstract}
Doped strong topological insulators are one of the most promising candidates to realize a fully gapped three-dimensional topological superconductor (TSC). In this letter, we revisit this system and reveal a possibility for higher-order topology which was  previously missed. We find that over a finite-range of doping, the Fu-Berg superconducting pairing can give rise to both Majorana surface states, and nodal hinge states. Interestingly, we observe the coexistence of surface and hinge modes in the superconducting state only when there are both bulk and surface Fermi-surfaces in the normal state. Also, we find that the hinge modes can appear for normal states consisting of doped strong or weak topological insulators. In summary, this work may allow for the discovery of superconducting hinge modes in a well explored class of materials, i.e., doped strong or weak topological insulators.
\end{abstract}
\maketitle

\emph{\blue{Introduction}}.-- Since its discovery, the notion of topological phases has been extended to almost all aspects of condensed matter physics and related fields, such as photonics and cold-atomic gasses. The initial topological AZ classification table has been extended multiple times and now includes crystalline phases and even gapless phases \cite{reviewTITSC,Chiu2016,reviewweyl}. One of the most recent additions are the so-called $nth$-order topological phases which possess gapless states on boundaries having co-dimension $d_c = n$ \cite{Benalcazar2017-1,Benalcazar2017-2,Song2017,Langbehn2017,Schindler2018-1,Schindler2018-2,Peterson2018,Imhof2018,Noh2018,ghorashi2019vortex,ghorashihowsm}. Hybrid-order topological phases where $1st$ and $2nd$ order boundary states coexist have also been discovered \cite{Ghorashihosc2019,kooi2019hybrid,ghorashihowsm}. Despite the many experimentally verifiable material candidates for topological insulators and topological semimetals, the search for topological superconductors (TSC) has proven to be more challenging. The challenge lies in finding a single material in which the band structure and interaction-driven superconductivity conspire to form a topological phase. Finding a higher-order generalization of a TSCs usually requires additional bandstructure features and/or exotic pairing, further complicating the search for experimental candidates \cite{Wang2018-TH}.

\begin{figure}[t!]
    \includegraphics[width=0.51\textwidth]{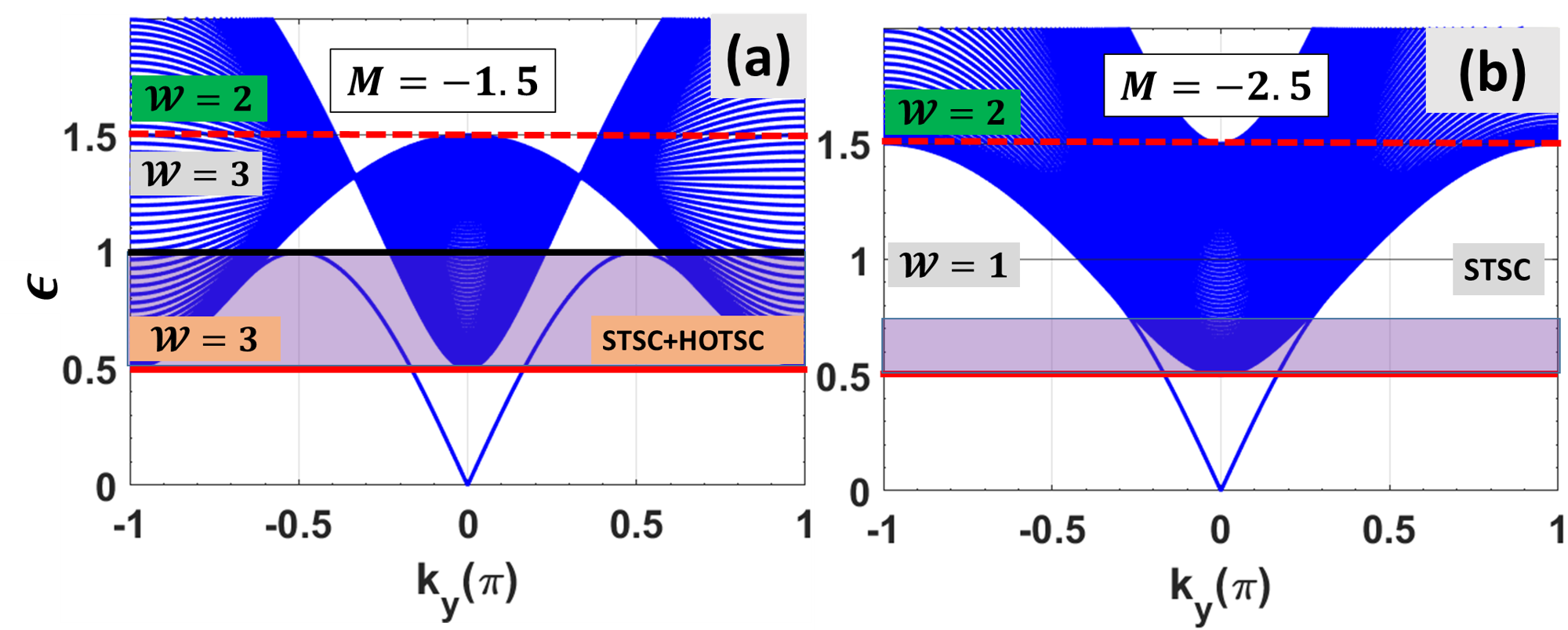}
    \caption{Summary of results. (a) For TIs described by $H_{TI}$ with $|M|<2$, there are three distinct topological superconducting phases achievable by doping if we assume the bulk pairing takes the Fu-Berg pairing\cite{Fu2010}. As an example, for $M=-1.5,$ when increasing $\mu$ from zero the system evolves to a strong TSC (STSC) (red solid line) with a winding number $\mathcal{W}=3$. By further increasing $\mu$ into the region where bulk and surface Fermi-surfaces coexist in the normal state (shaded purple region), we find a STSC that coexists with a HOTSC with nodal hinge states (labelled by $STSC + HOTSC$). Upon further increasing $\mu$, the hinge states vanish (black solid line), realizing a STSC. Finally another TSC appears having $\mathcal{W}=2$ (red dashed line) as $\mu$ increases further. In contrast, for (b) $|M|>2$ we find STSC phases but we do not find a HOTSC phase and corresponding hinge states.}
    \label{fig:adpic}
\end{figure}

\indent In this Letter, we revisit a family of promising solid state candidates for $1st$ order 3D TSCs: doped 3D topological insulators with time-reversal symmetry \cite{SCTI1,Fu2010,SCTI2,PhysRevLett.107.217001,PhysRevB.85.180509,Satoreview_2017}. We show that, remarkably, a system exhibiting the Fu-Berg inter-orbital singlet pairing\cite{Fu2010} can be tuned into a hybrid-order TSC (HyTSC) phase with coexisting surface and hinge states by doping such that both bulk and surface Fermi surfaces are present in the normal state. The surface Majorana cones are protected by time-reversal symmetry, while the hinge modes are protected by mirror symmetries, so they can be manipulated independently. 
Furthermore we find that the HyTSC state can be achieved from a strong or weak TI normal state. We demonstrate these phenomena using a conventional TI model with cubic symmetry. We then briefly show that the hinge states persist even in the absence of cubic symmetry (e.g., in the presence of hexagonal warping) \cite{warping1}, 
which might be relevant to realistic experimental contexts. Our main results are summarized in Fig. \ref{fig:adpic}.  

\blue{\emph{Model}}.-- We start from a model of a superconducting TI with the BdG Hamiltonian $H(\vex{k})=\sum \Psi^{\dagger} h(k) \Psi$ \cite{Fu2010},
where
\begin{align}\label{hscti}
h(\vex{k})=\Bigg[\begin{array}{cc}
   H_{TI}(\vex{k})-\mu  & -i\Delta \\
    i\Delta^{\dagger} & -H^T_{TI}(-\vex{k})+\mu
\end{array}\Bigg]
\end{align}
and
\begin{align}\label{eqTI}
    H_{TI}(\vex{k})=&\,\left(M+t_0\sum_i \cos (k_i)\right)\kappa_z\sigma_0\cr
+&\,t_1\sum_i\sin (k_i)\kappa_x\sigma_i,
\end{align} where $\kappa_a, \sigma_b$ $a,b=x,y,z$ are Pauli matrices in orbital and spin space, respectively, and $\kappa_0, \sigma_0$ are identity matrices.
In the following we set $t_0=t_1=1$. For the ranges $1<|M|<3$, $0<|M|<1,$ and $|M|>3$, $H_{TI}$ represents a time-reversal invariant insulator with strong, weak, and trivial topology, respectively. The pairing term is chosen to be the Fu-Berg pairing\cite{Fu2010}: $\Delta=\delta_0\kappa^1\sigma^2$ where $\delta_0$ is the pairing amplitude.   It has been shown that this inter-orbital, odd-parity pairing $\Delta$ is a relevant pairing instability in superconducting doped TIs \cite{Fu2010}. The Hamiltonian \eqref{hscti} preserves both time-reversal and inversion symmetries, as well as fourfold rotation, $C^z_4$. In the following we will discuss the HyTSC phase in this system, which exhibits coexisting Majorana surface states and mirror-protected nodal hinge states.

\begin{figure}[t!]
\includegraphics[width=0.49\textwidth]{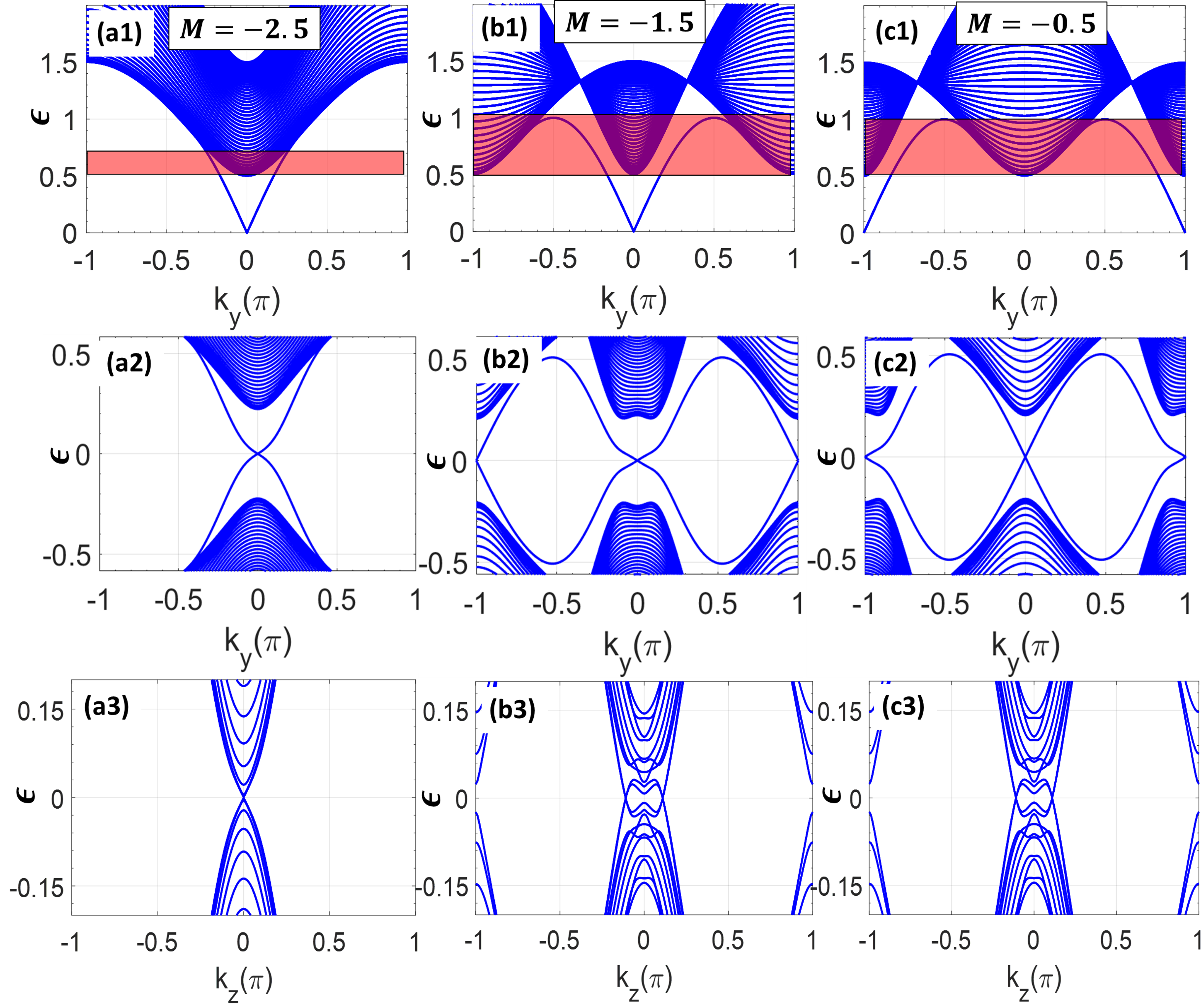}
\caption{(a1-c1) Spectrum with open boundaries in the $z$-direction (at $k_x=0$) of insulating phases of $H_{TI}$ in Eq.~\eqref{eqTI} with $M=-2.5$ (strong TI),$-1.5$ (strong TI),$-0.5$ (weak TI), respectively. (a2-c2) and (a3-c3) show surface (open in $z$-direction) and hinge (open in $x$ and $y$-directions) spectra of superconducting phases corresponding to (a1-c1), respectively. The parameters $\delta_0=0.4,\,\mu=0.6$ are used for all superconducting spectra.}
\label{Mcompare_surfhinge}
\end{figure}

\blue{\emph{Normal phase}}.-- While the precise value of $|M|$ in each of the ranges listed above has no bearing on the topological properties of the \emph{insulating} state, it can, however, have a notable effect on the nature of the Fermi surfaces at a given Fermi level. The different Fermi surface structures are correlated with the resulting TSC phases. For illustration we show representative examples of the normal phase spectrum of $H_{TI}$ for  values of $M=-2.5,\,-1.5,\,-0.5$ in Fig.~\ref{Mcompare_surfhinge}(a1-c1). These spectra are calculated for open boundaries in $z$, but periodic in $x,y.$ We show a particular $k_x=0$ slice where one can see the bulk gap (which is the same size for all three values of $M$) and midgap surface states. The values $M=-2.5, -1.5$ are strong TIs, while $M=-0.5$ is a weak TI which has an even number of surface Dirac cones (only one is visible at $k_x=0$ because the other is centered around $k_x=\pi$ ).

The first two values of $M$, despite being in the same topological phase, have qualitative differences in their surface and bulk Fermi surfaces that turn out to be important for the characterization of the TSC phase. To illustrate this, let us focus on the red bars marked in each of Fig.~\ref{Mcompare_surfhinge}(a1-c1). These denote a window of energy in which the surface and bulk states coexist (which we denote the surf-bulk region), and which varies as $M$ changes. The coexistence region is generic as there are only two fine-tuned values of $M=0,-2$ (see Fig.~\ref{M2and35_surfhinge}(a1)) where the system is insulating and the surf-bulk region vanishes. We also note that the closer the value of $M$ is to the weak-to-strong TI phase transition at $M=-1$, the wider the surf-bulk region becomes.  We find that as long as there is a finite surf-bulk region, the weak or strong topological nature of the normal phase does not have a major effect on the appearance of hinge-modes, as we discuss below. This, and the fact the the phenomena we examine survive to the weak pairing limit, are an indication that the physics is driven by the Fermi surfaces and not the topology of the occupied bands. We also note that the evolution of the surface Fermi surfaces as one raises the chemical potential can be different within the same topological phase, e.g., $M=-2.5, -1.5,$ in Fig.~\ref{Mcompare_surfhinge}(a1,b1). In the case of $M=-2.5$ the surface Dirac cone at the $\Gamma$ point evolves into bulk Fermi surfaces encircling the $\Gamma$ point, while for $M=-1.5$, the surface Dirac point evolves to bulk Fermi surfaces centered at $(k_x,k_y)=(0,\pi)$ and $(\pi,0)$. 

\blue{\emph{Superconducting phase}}.-- Now let us consider the BdG quasiparticle spectrum for this system. The bulk (1st-order) topology protected by time-reversal symmetry can be characterized using the 3D winding number\cite{Schnyder2008,Satoreview_2017}:
\begin{align}
    \mathcal{W}=\frac{1}{48\pi^2}\int d\vex{k}\epsilon_{ijk}\large[\mathcal{S}(h^{-1}\partial_i h)(h^{-1}\partial_j h)(h^{-1}\partial_k h)\large],
\end{align}
where $\mathcal{S}$ is a chiral symmetry that acts as $\mathcal{S}h(\vex{k})\mathcal{S}^{-1}=-h(\vex{k})$. For the DIII class, $\mathcal{W}$ takes integer values and guarantees the existence of the surface Majorana cones when it is non-vanishing\cite{Schnyder2008,Volovik_2009,Satoreview_2017}. In our model, at a representative value of $\delta_0=0.4,$ for example, the bulk superconducting spectrum undergoes topological phase transitions as a function of $\mu.$
In Fig.~\ref{fig:adpic}, we show $\mathcal{W}$ over a range of $\mu$ for two representative values of $M=-2.5, -1.5$ (see \cite{sm} for more detail). Both values of $M$ have a topological phase transition at $\mu_{c1}=E_g-\delta_0/2=0.3$, where $E_g$ is the insulating gap.  For $|M|<2$ the winding number transitions from zero to $\mathcal{W}=3$ at $\mu_{c1}$ while for $|M|>2$ it transitions to $\mathcal{W}=1$. In the ultra weak pairing limit as $\delta_0\to 0$ this difference can be ascribed to the number of (spin-degenerate) Fermi surfaces affected by the pairing, i.e., for $|M|<2$ there are three closed Fermi surfaces and each one surrounds one of the $X,Y$ and $Z$ points in the BZ, while for $|M|>2$ there is a single closed Fermi surface surrounding the $\Gamma$-point.
The conventional bulk-boundary correspondence indicates that a higher-winding number signals higher numbers of stable surface states. Therefore, even at the level of $1st$-order topology the doped $|M|<2$ and $|M|>2$ insulators generate different superconducting topological phases, even though their corresponding normal phases are topologically equivalent. Since our model has $C^{z}_4$-symmetry, one can also obtain the winding number mod 4 using symmetry indicators. For example, the occupied bands when $M=-2.5 (M=-1.5)$ have one (three) pair(s) of negative inversion eigenvalues having $C^z_4$ eigenvalues of $e^{\pm i\pi/4}$ at the four $C^z_4$-symmetric momenta. Using the results of \cite{Ono_2020}, these results are consistent with our integral calculations.

\begin{figure*}[hbt!]
\includegraphics[width=1\textwidth]{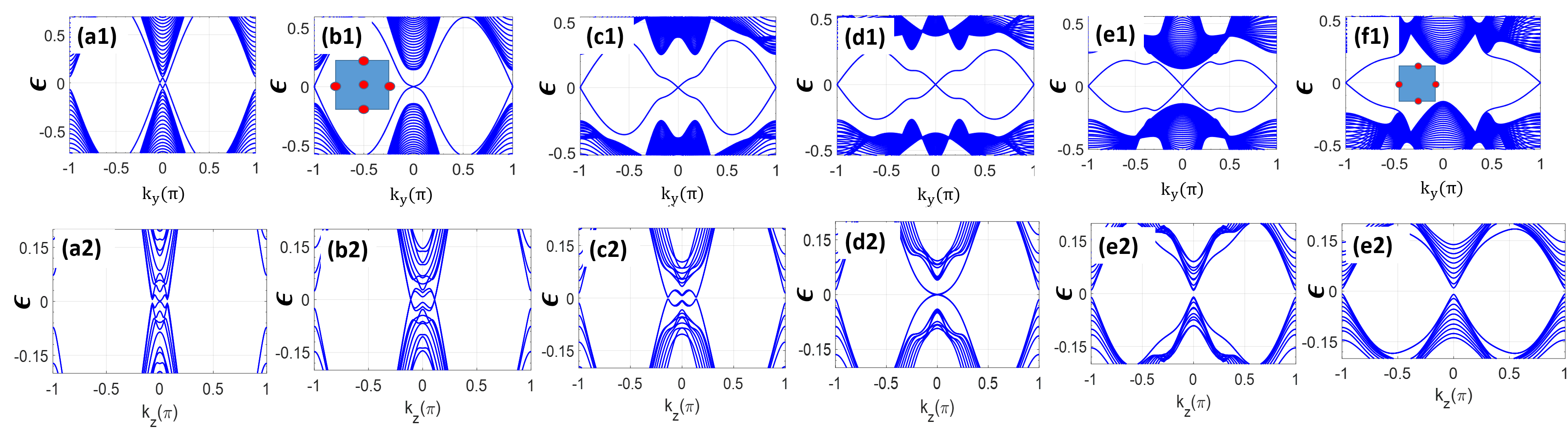}
\caption{The evolution of surface (a1-f1) and hinge (a2-f2) states in the superconducting strong TI versus chemical potential $\mu$. The values of $\mu=0.4,0.6,0.8,1,1.3,1.6$, are used for (a-f), respectively.
Insets in (b1) and (f1) show the the 2d arrangement of surface nodes. The parameters $M=-1.5,\,\delta_0=0.4$ are used throughout.}
\label{fig:surfhingevsMu_Mn1_5}
\end{figure*}
Now, let us examine the surface states in the superconducting phases more closely. In Fig.~\ref{Mcompare_surfhinge}(a2-c2) we show the surface corresponding to the Hamiltonian Eq.~\eqref{hscti} with open boundaries in the $z$-direction, for $M=-2.5,\, -1.5,\, -0.5$ at fixed values of $\delta_0=0.4, \mu=0.6$; the other surfaces behave similarly due to the crystal symmetries. Note that the chemical potential is within the surf-bulk region denoted by the red bars in Fig.~\ref{Mcompare_surfhinge}(a1-c1). For $M=-2.5$, there is only one Majorana cone located at  $k_x=k_y=0$ in the reduced BZ, while the phase with $M=-1.5$ hosts three Majorana cones: one at $k_x=k_y=0$ and two others at $(k_x, k_y)=(0, \pi),  (\pi,0)$, in agreement with the associated bulk invariants of $\mathcal{W}=1$ and $3$, respectively. In the $\mathcal{W}=3$ phase there are two gapless points on the $k_x=0$ and $k_y=0$ lines; along these lines, the surface states attach to each other and are \emph{free-floating}, i.e., they are detached from the bulk states. This surface-state connectivity requires $|\mathcal{W}|>1.$

Remarkably, when cutting the surface again to form hinges the phase with $\mathcal{W}=3$ has nodal hinge states when the chemical potential is in the surf-bulk region.  
To see this explicitly, in Fig.~\ref{fig:surfhingevsMu_Mn1_5} we have plotted the evolution of the superconducting surface and hinge states versus $\mu$ for a TI with $M=-1.5$. We find that the hinge states appear only when $0.5<\mu<1$,  which corresponds to the surf-bulk region in the normal phase, (c.f., Fig.~\ref{Mcompare_surfhinge}(b1)), but not the full span of the $\mathcal{W}=3$ phase. We also note that, despite having its own surf-bulk coexistence region, the $\mathcal{W}=1$ phase does not exhibit such hinge states.

We can understand more about the nature and protection of the hinge states by considering the low-energy Hamiltonian on a single hinge. Consider the critical point of the hinge states at $M=-1.5, \delta_0=0.4,$ and $\mu=1.0$ as shown in Fig. \ref{fig:surfhingevsMu_Mn1_5}(d2). At this critical point the energy spectrum of the modes on a single hinge consists of two parabolas that touch at the high symmetry point $k_z=0$. The critical Hamiltonian on a single hinge is hence given by
\begin{equation}
    H_{hinge}(k_z)=k_z^2\sigma^z.
\end{equation} This low-energy Hamiltonian has a particle-hole symmetry implemented by $C=\sigma^x$, such that $CH_{hinge}(k_x)C^{-1}=-H_{hinge}^{T}(-k_z).$ We also want the Hamiltonian to have mirror symmetry $M_z$, and
there are two choices $M_z=\mathbb{I}$ or $M_z=\sigma^z.$ Both choices lead to $M_z H_{1D}(k_z)M_z^{-1}=H_{1D}(-k_z).$ However, the bulk Hamiltonian has operators that satisfy $[C,M_z]=0$ so we choose $M_z=\mathbb{I}$ (also if we chose $M_z=\sigma^z$ the mirror symmetric spectrum would not match the numerical results).\footnote{We note that this low-energy model has an effective time-reversal symmetry such that $\mathcal{T}^2=+1,$ while the original Hamiltonian has $\mathcal{T}^2=-1.$ When $\mu\neq 1$ the states at $k_z=0$ combine with other bulk and surface states to restore the full $\mathcal{T}^2=-1$ symmetry.}

The generic Hamiltonian of the hinge states satisfying $C$ and $M_z$ is given by
\begin{equation}
     H^{\mathbb{I}}_{1D}(k_z)=k_z^2\sigma^z+(\mu-1)\sigma^z,
\end{equation}
to second order in $k_z$.
This Hamiltonian has energy bands $E(k_z)=\pm(k_z^2+\mu-1).$ If $\mu>1$ the two parabolas lift off each other and generate a gapped phase as seen in numerics. If $\mu<1$ the base of the upper parabola moves below the peak of the lower parabola leaving gapless, linear crossing points at $k_z^{\ast}=\pm \sqrt{|\mu-1|}.$

We expand the Hamiltonian around the two $k_z^{\ast}$ to find the four-band Hamiltonian for the two valleys:
\begin{equation}
H_{V}(k_z)=2\sqrt{|\mu-1|}k_z \tau^z\sigma^z,
\end{equation}
where $\tau^z$ acts in the valley space and $k_z$ represents the deviation away from $\pm k^{\ast}_z$ in each block respectively. Both mirror and particle-hole exchange the valleys since they flip the sign of the $k_z$ momentum. Hence we would expect that in the valley Hamiltonian $C^V=\tau^x\sigma^x$ and $M^{V}_{z}=\tau^x\mathbb{I}.$ One can check that our Hamiltonian in the valley representation still has both $C$ and $M_z$ symmetry and that $[C^V,M^V_z]=0$ still holds.

To check the stability of the gapless hinge modes we enumerate all possible gapping terms. The full list is $(1) \tau^z\sigma^x,$ $(2) \tau^z\sigma^y,$ $(3) \mathbb{I}\sigma^x,$ $(4) \mathbb{I}\sigma^y,$ $(5) \tau^x\sigma^z,$ $(6) \tau^x\mathbb{I},$ $ (7) \tau^{y}\sigma^z,$ $ (8) \tau^{y}\mathbb{I}.$
Each mass term breaks a symmetry, specifically: mass (1) and (2) break mirror, mass (3) and (4) break particle-hole, mass (5) breaks translation, mass (6) breaks translation and particle hole, mass (7) breaks translation and mirror, and mass (8) breaks particle hole, mirror, and translation. Let us ignore translation for a moment. We find that if we do not have mirror we can add for example mass (1). If we do not have particle-hole we can add mass (3) so they are both required for protection. If we have both symmetries then all masses except (5) are forbidden, and we can forbid (5) by assuming translation symmetry along the hinge, as is implicitly required to protect nodal points in general.

Now let us consider the classification of the nodal points. Suppose we make two identical copies of our Hamiltonian
\begin{equation}
H^{double}_{V}(k_z)=2\sqrt{|\mu-1|}k_z \mathbb{I}\tau^z\sigma^z.  \end{equation} Because the copies are identical, the symmetries are $C^V=\mathbb{I}\tau^{x}\sigma^{x}$ and $M_z^V=\mathbb{I}\tau^x\mathbb{I}.$ Now consider mass (3) which broke only particle hole symmetry. For the doubled Hamiltonian we can have the mass $(3')\mu^y\mathbb{I}\sigma^x$, where $\mu^a$ are Pauli matrices in the double copy space. Since we have added an imaginary matrix as compared to mass (3), this mass term will now preserve particle-hole. Adding the extra $\mu^y$ factor does not affect translation or mirror so this mass term now obeys all of the symmetries. Hence the stability classification of the nodal points is $\mathbb{Z}_2$ since two identical copies can be gapped while preserving all of the symmetries. Thus, while one hinge is stable in the presence of the symmetries mentioned, if two hinges are coupled then the modes can be gapped out.

We now comment on two specific cases which are particularly instructive. Fig.~\ref{M2and35_surfhinge} shows the surface (normal and superconducting) and hinge (superconducting) plots for $M=-2$ and $M=-3.5$. The former is a strong TI, while the latter is a trivial insulator, and both cases lack a surf-bulk region. We make two remarks: (i) while the emergence of $1st$-order topological superconductivity does not require non-trivial bulk topology of the normal phase, we find hinge states in this system only when the normal phase possesses surface states generated by either weak or strong topology; and (ii) if we pick a value of $|M|\sim 2$ and fix $\mu$ such that hinge states are generated, then tuning through $|M|=2$ while keeping $\mu$ fixed will drive a transition between the HyTSC phase and a $1st$-order topological superconductor since the normal state for $|M|=2$ does not exhibit coexisting surface and bulk Fermi-surfaces for any value of $\mu.$
\begin{figure}[htb!]
\includegraphics[width=0.48\textwidth]{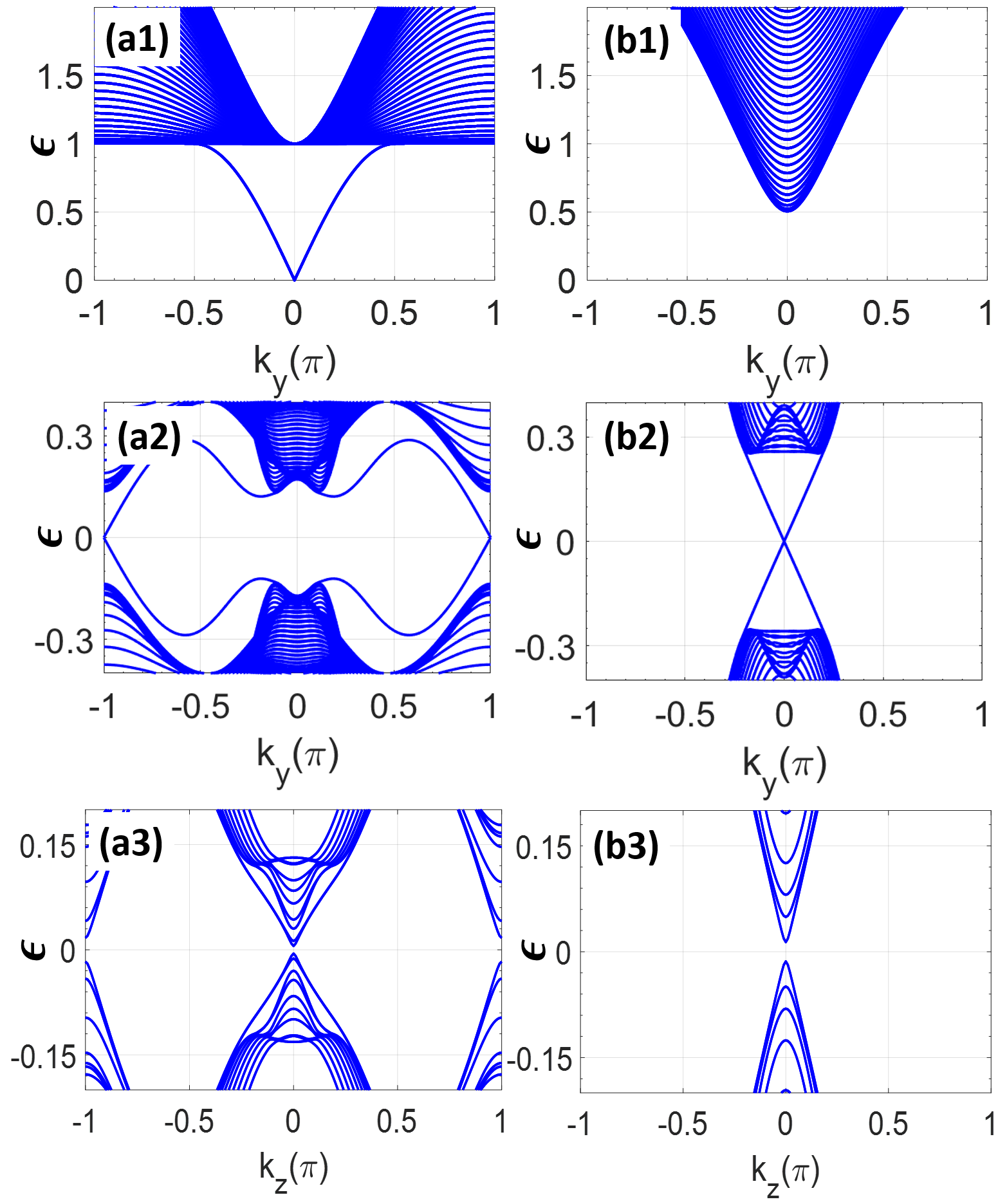}
\caption{Surface states of the normal phase, and surface and hinge states of the superconducting phase, for $M=-2$ (a1-a3) and $M=-3.5$ (b1-b3), respectively. The parameters $\delta_0=0.4$, $\mu=1.2$(a2,a3) and $\delta_0=0.4, \mu=0.8$ (b2,b3) are used.}
\label{M2and35_surfhinge}
\end{figure}

\blue{\emph{Discussion}}.--We now make a few final remarks. First, as we have shown, the higher-order gapless hinge states appear for both weak and strong TIs. Hence, our finding facilitates the experimental realization of hinge states by broadening the range of possible material candidates. Second, 
in many realistic TI materials, such as Bi$_2$Te$_3$ the surface states are subject to hexagonal warping\cite{warping1}. In the supplement\cite{sm} we investigate the effect of a particular type of hexagonal warping and find that hinge states can still appear, though the deformation caused by the warping affects the parameter ranges and spatial location of the hinge states compared to the case with $C^z_4$ symmetry studied here. A more detailed analysis of the effects of hexagonal warping is left for future work. As the last remark, we note that recently, there have been many developments in characterization of TSCs via symmetry indicators \cite{Ono_2020,SI3,SI1,huang2020faithful}. A particular example is a recent work \cite{kooi2019hybrid}, which proposed a 3D hybrid-order TSC in presence of $C^z_4$ symmetry using a more complex tight-binding model and pairing. However, the invariants introduced there can not be applied to our system as the hinge states we find are more akin to a 1D nodal superconductors protected by mirror symmetry and do not appear at high-symmetry momenta.

Finally, we emphasize that the physics discussed in this work opens up a new avenue toward the experimental discovery of HOTSCs using well-studied material candidates, in either weak and strong TIs phases, and calls for further careful experimental studies in these systems, especially in scenarios where the chemical potential intersects both bulk states and topological surface states. \\

\emph{Acknowledgement.}-- We thank Masatoshi Sato for useful discussion. S.A.A.G acknowledges support from the Air Force Office of Scientific Research (Grant No. FA9550-20-1-0260). ER acknowledges support from DOE-BES, Grant No. DE-SC0022245. T.L.H thanks ARO MURI W911NF2020166 for support.
J.C. is partially supported by the Alfred P. Sloan Foundation through a Sloan Research Fellowship and acknowledges the support of the Flatiron Institute, a division of the Simons Foundation.


%

\newpage \clearpage

\onecolumngrid

\begin{center}
	{\large
	SUPPLEMENTAL MATERIAL: Theoretical Discovery of Higher-Order Topology in Superconducting Doped Topological Insulators
	\vspace{4pt}
	\\
	
	Sayed Ali Akbar Ghorashi, Jennifer Cano, Enrico Rossi, Taylor L. Hughes}
\end{center}

\section{I.\, Hexagonal warping}
\begin{figure}[htb!]
    \centering
    \includegraphics[width=0.8\textwidth]{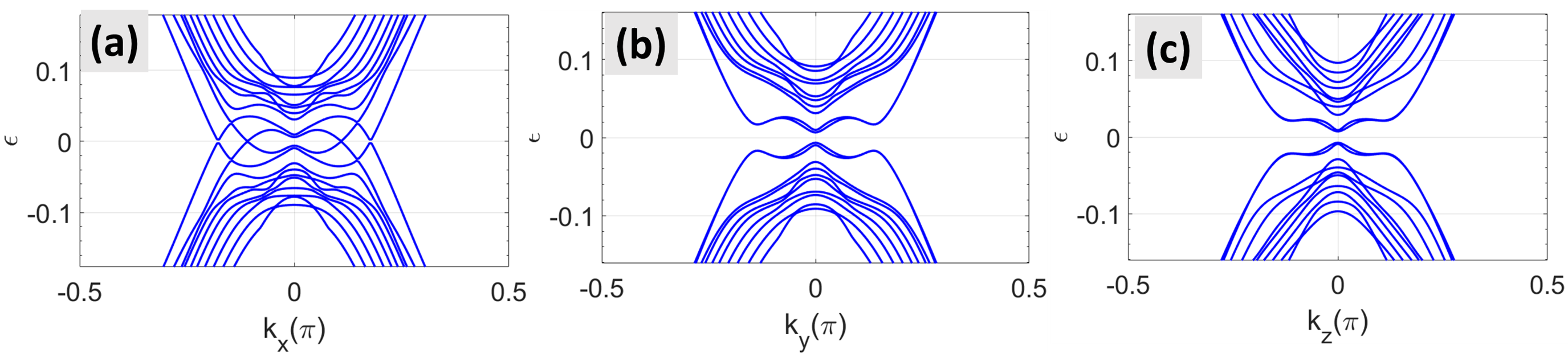}
    \caption{The hinge spectrum of superconducting TI along the (a) $x$, (b), $y$ and (c) $z$ directions. We used $M=-1.5,\,\delta_0=0.4,\,\mu=0.8$ and $R_1=0.1$.}
    \label{fig:warping}
\end{figure}

Here, we break the cubic symmetry by addition of a term of ($\mathcal{O}(\vex{k}^3)$) which in the hexagonal lattices represent the the effect of hexagonal warping. This will show that the cubic symmetry of the model is not crucial and hinge states can appear in models with reduced symmetries. We use the following higher-order term which break in-plane rotation symmetry down to the threefold rotation symmetry,
\begin{align}
    H_{hex}=-\frac{R_1}{2}\big(k^3_{+}+k^3_{-}\big)\kappa_y\sigma_0
\end{align}
We convert it to a lattice representation as follows:
\begin{align}
    H_{hex}=-R_1\big(-4\sin(k_x)-\sin(2k_x)+6\sin(k_x)\cos(k_y)\big)\kappa_y\sigma_0
\end{align}
The rotational symmetry breaks for any non-zero values of $R_{1}$. However, $R_1$ breaks also both the mirror symmetries along $y,z$ but keeps the $M_x$ intact. As a result we find that hexagonal warping gaps out the $y,z-$hinges but the $x-$hinge remains gapless and only splits the two eight-fold nodes to four four-fold nodes. Fig.~\ref{fig:warping} shows the $x,y,z$-hinge states of superconducting phase for weak hexagonal warping $R_1$.

\section{II.\, Table of all allowed perturbations}
Here we list table of all trilinear perturbations $\tau^{i}\kappa^j\sigma^k$ allowed by particle-hole symmetry. This table is produced for the following set of parameters: $M=-1.5,\,\delta_0=0.4,\,\mu=0.8$ (as used in Fig.~3(c1,c2)).

\begin{table}[H]
\begin{tabular}{|c||c|c|c|c|c|c|c|c|c|c|}\hline
     & $M_{x}=IZX$ & $M_{y}=ZZY$ & $M_{z}=IZZ$ & $C_{2x}=ZIX$ & $C_{2y}=IIY$ & $C_{2z}=ZIZ$ & $T=IIY\kappa$ & $\mathbb{I}=ZZI$ & Surf & Hinge \\\hline\hline
     IIY & 0 & 1 & 0 & 0 & 1 & 0 & 0 & 1 & (1,0,1) & (1,1,1) \\\hline
     IXY & 1 & 0 & 1 & 0 & 1 & 0 & 0 & 0 & (1,1,1) & (1,1,1) \\\hline
     IYI & 0 & 0 & 0 & 1 & 1 & 1 & 0 & 0 & (0,0,0) & (1,1,1) \\\hline
     IYX & 0 & 1 & 1 & 1 & 0 & 0 & 1 & 0 & (1,1,1) & (1,1,1) \\\hline
     IYZ & 1 & 1 & 0 & 0 & 0 & 1 & 1 & 0 & (1,1,1) & (1,1,1) \\\hline
     IZY & 0 & 1 & 0 & 0 & 1 & 0 & 0 & 1 & (1,0,1) & (0,1,0) \\\hline
     XIY(mass) & 0 & 0 & 0 & 1 & 1 & 1 & 0 & 0 & (0,0,0) & (0,0,0) \\\hline
     XXY & 1 & 1 & 1 & 1 & 1 & 1 & 0 & 1 & (1,1,1) & (1,1,1) \\\hline
     XYI & 0 & 1 & 0 & 0 & 1 & 0 & 0 & 1 & (1,0,1) & (1,1,1) \\\hline
     XYX & 0 & 0 & 1 & 0 & 0 & 1 & 1 & 1 & (1,1,1) & (0,0,1) \\\hline
     XYZ & 1 & 0 & 0 & 1 & 0 & 0 & 1 & 1 & (1,1,1) & (1,0,0) \\\hline
     XZY & 0 & 0 & 0 & 1 & 1 & 1 & 0 & 0 & (0,0,0) & (1,1,1) \\\hline
     YIY & 0 & 0 & 0 & 1 & 1 & 1 & 1 & 0 & (1,1,1) & (1,1,1) \\\hline
     YXY & 1 & 1 & 1 & 1 & 1 & 1 & 1 & 1 & (1,1,1) & (1,1,1) \\\hline
     YYI & 0 & 1 & 0 & 0 & 1 & 0 & 1 & 1 & (1,1,1) & (0,1,0) \\\hline
     YYX & 0 & 0 & 1 & 0 & 0 & 1 & 0 & 1 & (1,1,0) & (1,1,1) \\\hline
     YYZ & 1 & 0 & 0 & 1 & 0 & 0 & 0 & 1 & (0,1,1) & (1,1,1) \\\hline
     YZY & 0 & 0 & 0 & 1 & 1 & 1 & 1 & 0 & (1,1,1) & (1,1,1) \\\hline
     ZII & 1 & 1 & 1 & 1 & 1 & 1 & 1 & 1 & (1,1,1) & (1,1,1) \\\hline
     ZIX & 1 & 0 & 0 & 1 & 0 & 0 & 0 & 1 & (0,1,1) & (1,1,1) \\\hline
     ZIZ & 0 & 0 & 1 & 0 & 0 & 1 & 0 & 1 & (1,1,0) & (1,1,1) \\\hline
     ZXI & 0 & 0 & 0 & 1 & 1 & 1 & 1 & 0 & (1,1,1) & (0,0,0) \\\hline
     ZXX & 0 & 1 & 1 & 1 & 0 & 0 & 0 & 0 & (1,1,1) & (1,1,1) \\\hline
     ZXZ & 1 & 1 & 0 & 0 & 0 & 1 & 0 & 0 & (1,1,1) & (1,1,1) \\\hline
     ZYY & 1 & 0 & 1 & 0 & 1 & 0 & 1 & 0 & (1,1,1) & (1,1,1) \\\hline
     ZZI & 1 & 1 & 1 & 1 & 1 & 1 & 1 & 1 & (1,1,1) & (1,1,1) \\\hline
     ZZX & 1 & 0 & 0 & 1 & 0 & 0 & 0 & 1 & (0,1,1) & (1,0,0) \\\hline
     ZZZ & 0 & 0 & 1 & 0 & 0 & 1 & 0 & 1 & (1,1,0) & (0,0,1) \\\hline
\end{tabular}
\caption{``0'' and ``1'' denote the absence or existence of the indicated symmetry.
For surface and hinge, it is shown as (x,y,z) vector of the corresponding cuts with ``0'' and ``1'' showing gapped and gapless boundaries respectively. For example (1,0,1) for surface means surfaces perpendicular to the $x,z$ remain gapless but the $z$-surface is gapped. For the hinge (1,0,1) means the hinges parallel to $x,z$ and $y$ are gapless and gapped respectively. The (mass) indicates the perturbation which gaps out all the surface and hinge states. }
\end{table}

\section{III.\, Superconducting Weak TI}
\begin{figure}[H]
    \centering
    \includegraphics[width=1.05\textwidth]{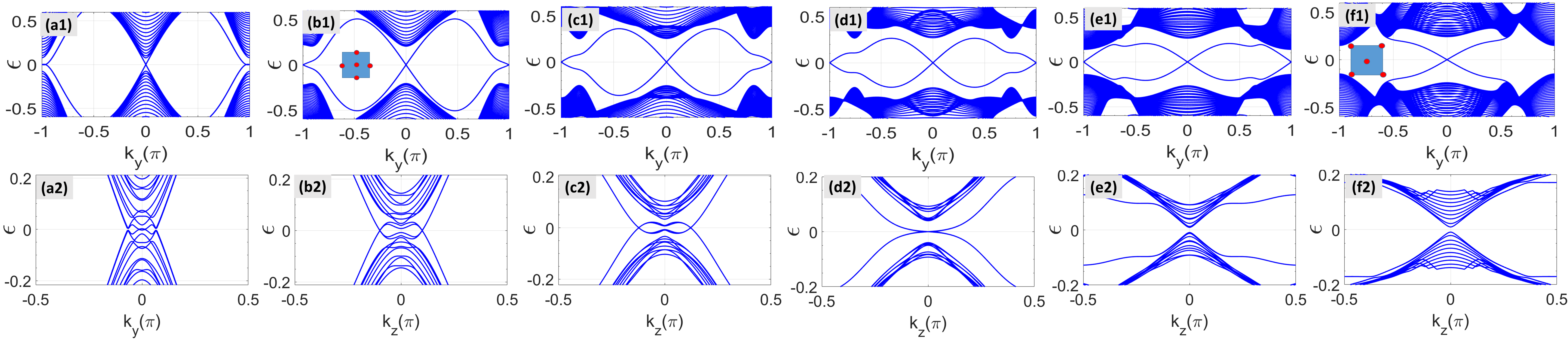}
    \caption{The evolution of surface (a1-f1) and hinges (a2-f2) states in Superconducting weak TI versus chemical doping $\mu$. $\mu=0.4,0.6,0.8,1,1.3,1.6$, is used for (a-f), respectively.The inset in (b1) and (f1) shows the the 2d arrangement of surface nodes. $M=-1.5,\,\delta_0=0.4$.}
    \label{fig:SCWTI}
\end{figure}

Here we show the corresponding plots of Fig.~3 of the main text for the case of weak topological insulators. This clearly shows that despite the fact that the normal phases of weak and strong TIs possess very different topology, the Majorana hinge modes can universally appear for both the superconducting weak and strong TIs. In fact there is no difference between the superconducting phases of weak ad strong TIs in terms of $1st$-order boundary and topology. Similar to the superconducting strong TI, the first order topology evolves from trivial to the $W=3$ and then $W=2$ topological phases. However, in the $W=2$ phase, the two Majorana surface cones in the superconducting WTI are positioned at the center and four corners (inset in \ref{fig:SCWTI}(f1)) while for the superconducting STI they are located at the four sides of the BZ (inset in Fig.~3(f1) of the main text).

\end{document}